\documentclass[sigconf]{acmart}
\AtBeginDocument{%
  }
\usepackage{multirow}

\copyrightyear{2026}
\acmYear{2026}
\setcopyright{cc}
\setcctype{by-nc-nd}
\acmConference[CHI '26]{Proceedings of the 2026 CHI Conference on Human Factors in Computing Systems}{April 13--17, 2026}{Barcelona, Spain}
\acmBooktitle{Proceedings of the 2026 CHI Conference on Human Factors in Computing Systems (CHI '26), April 13--17, 2026, Barcelona, Spain}
\acmPrice{}
\acmDOI{10.1145/3772318.3790486}
\acmISBN{979-8-4007-2278-3/2026/04}

\usepackage{colortbl} 
\definecolor{mypink1}{RGB}{237, 251, 251}
\definecolor{mypink2}{RGB}{252, 237, 251}

\newcommand{\rowcol}{\rowcolor{mypink2}}
\newcommand{\rowwhi}{\rowcolor{white}}

\begin{document}

\title{\emph{Reflective Motion and a Physical Canvas}: Exploring Embodied Journaling in Virtual Reality}


\author{Michael Yin}
\orcid{0000-0003-1164-5229}
\affiliation{
  \institution{University of British Columbia}
  \city{Vancouver}
  \state{BC}
  \country{Canada}
  \postcode{V6T 1Z4}
}
\email{jiyin@cs.ubc.ca}

\author{Robert Xiao}
\orcid{0000-0003-4306-8825}
\affiliation{
  \institution{University of British Columbia}
  \city{Vancouver}
  \state{BC}
  \country{Canada}
  \postcode{V6T 1Z4}
}
\email{brx@cs.ubc.ca}

\author{Nadine Wagener}
\orcid{0000-0003-4572-4646}
\affiliation{
  \institution{OFFIS - Institute for Informatics}
  \city{Oldenburg}
  \country{Germany}
}
\email{nadine.wagener@offis.de}


\renewcommand{\shortauthors}{}

\begin{abstract}
In traditional journaling practices, authors express and process their thoughts by writing them down. We propose a somaesthetic-inspired alternative that uses the human body, rather than written words, as the medium of expression. We coin this \emph{embodied journaling}, as people's isolated body movements and spoken words become the canvas of reflection. We implemented embodied journaling in virtual reality and conducted a within-subject user study ($N=20$) to explore the emergent behaviours from the process, comparing its expressive and reflective qualities to those of written journaling. When writing-based norms and affordances were absent, we found that participants defaulted towards unfiltered emotional expression, often forgoing words altogether. Rather, subconscious body motion and paralinguistic acoustic qualities unveiled deeper, sometimes hidden feelings, prompting reflection that happens after emotional expression rather than during it. We discuss both the capabilities and pitfalls of embodied journaling, ultimately challenging the idea that reflection culminates in linguistic reasoning. 
\end{abstract}



\begin{teaserfigure}
  \centering
  \includegraphics[width=0.9\textwidth]{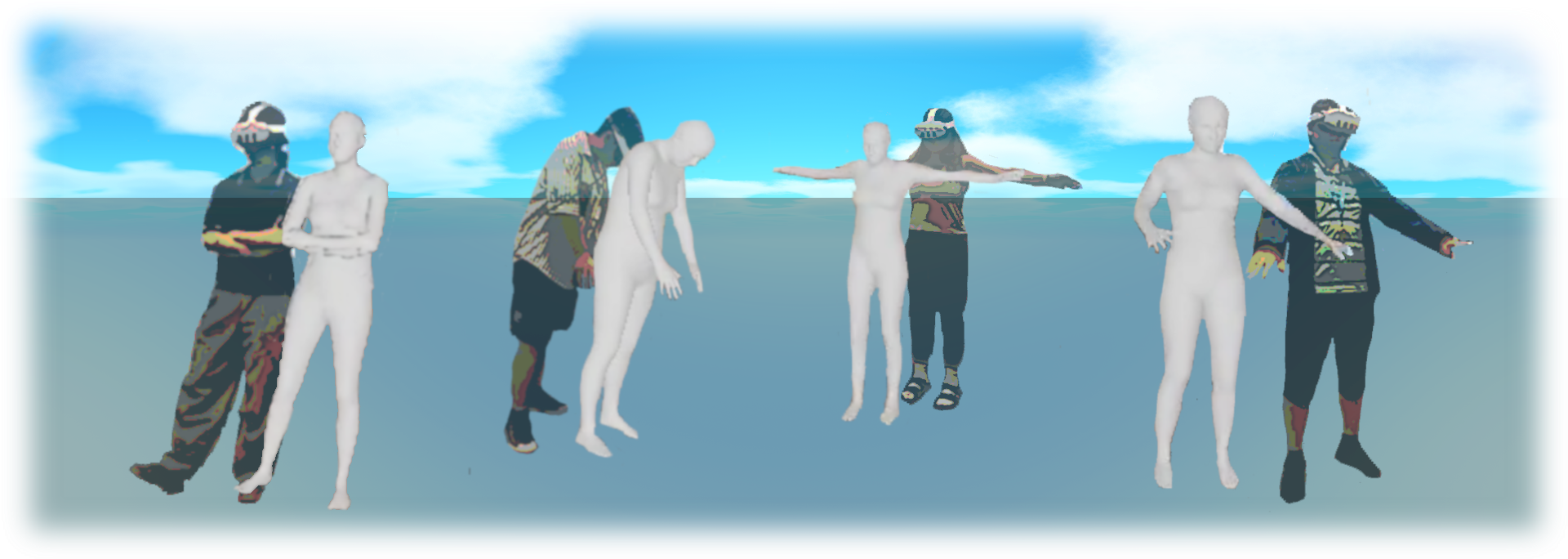}
  \caption{We study \emph{embodied journaling}, which uses the human body --- and its capabilities to move and speak --- for reflection. When people represent their emotions, what does it mean to cross their arms, slouch, speak, and pace around? Using VR as the medium to explore this experience, we study how people express, process, and reflect on a lived experience using their bodies.}
  \Description{In this picture, we see 4 pairs of humans and avatars side-by-side in a virtual world. Each of the humans is doing something to express their emotion, e.g. crossed arms, slouching, gesturing, moving their hands, and their nondescript avatar pair is mimicking their motion.}
  \label{fig:teaser}
\end{teaserfigure}

\maketitle

\section{Introduction}

\begin{quote}
    ``\emph{Movement, not language, is our primary way of understanding and being in the world.}'' 
    
    --- Kristina Höök \cite{hookDesigningBodySomaesthetic2018}, summarizing Maxine Sheets-Johnstone \cite{sheets-johnstoneThinkingMovement1981}
\end{quote}

Höök's writing about somaesthetics emphasises the \emph{primacy of movement}, which posits that movement precedes language in communication \cite{hookDesigningBodySomaesthetic2018}. Under this theory, \emph{postkinetic language} helps describe and fill in meaning derived from \emph{prelinguistic movement}. Movement emerges naturally and adapts to the surrounding world, and Höök underscores how all living beings create meaning through their movements in the world \cite{hookDesigningBodySomaesthetic2018}. 

Yet, the primary way in which meaning-making processes have been studied in HCI is through language-based communication. Through a rich lineage of work on written expression, researchers have shown how putting feelings and experiences into words supports meaning-making and reflection \cite{cowanFacilitatingReflectiveJournalling2013a, hubbsPaperMirrorUnderstanding2005, pennebakerTheoriesTherapiesTaxpayers2004, gortnerBenefitsExpressiveWriting2006}. However, writing inherently has a level of \emph{filtering} and \emph{processing} that is different from other forms of raw self-expression, including speaking \cite{sawhneyAudioJournalingSelfreflection2018a} and, hypothetically, the aforementioned motion. 

In this work, we draw inspiration from somaesthetics (e.g. \cite{shustermanSomaestheticsDisciplinaryProposal1999, hookDesigningBodySomaesthetic2018}) and related frameworks such as the Self-Care Technology Process Model (SCTpm)~\cite{wagenerSelfCareTechnologyProcess2025}, which emphasise the importance of movement for introspection and reflection. Departing from purely written forms of expression, we study how the \emph{human body} and its capacity for motion and speech can be used to express and reflect on experiences and emotions. Prior work has considered motion in the form of \emph{embodied reflection} \cite{buttingsrudBodiesSkilledPerformance2021, schwenderCreatingFramingReflecting2025, botelhoReflectionMotionEmbodied2021} in philosophical and educational domains. We formalise these theoretical works into a structured process that supports entries of isolated embodiment that can be constructed and reviewed. We coin this process --- of using physical motion paired with one's voice to create reviewable ``journal'' entries --- as \emph{embodied journaling} (Figure \ref{fig:teaser}). Embodied journaling harks back to emotion regulation~\cite{ gross2015ERcurrentstate,kateri2020ER} and metacognition therapy~\cite{Capobianco2023metacognitive-therapy}. We hope that embodied journaling, similarly to expressive writing \cite{gortnerBenefitsExpressiveWriting2006, pennebakerTheoriesTherapiesTaxpayers2004}, can support managing negative emotions, aligning with other immersive interactive technologies aiming to elicit positive change~\cite{kitson2018positivechange}. 

To explore embodied journaling, we employ virtual reality (VR), a technology that enables controlled investigation of body awareness and behaviour~\cite{dollingerVirtualRealityMind2022}. VR supports \emph{embodiment}, allowing users to inhabit a second body \cite{freemanMyBodyMy2020b} and providing a strong sense of self-location, agency, and body ownership \cite{kilteniSenseEmbodimentVirtual2012c}. Taking advantage of such affordances, VR has been used in prior research to convert abstract aesthetics into reflection and emotion regulation through, for instance, performing mundane tasks~\cite{raschMindMansionExploring2024, jiangPOVRDesigningImmersive2024}, interacting with symbolic representations of negative emotions~\cite{wagenerMoodShaper2024, bahngReflexiveVRStorytelling2020, wolfeImpactsDesignElements2022}, creating creative expressions~\cite{wagenerSelVReflectGuidedVR2023a, bahngDesigningImmersiveStories2023a}, or curating memories~\cite{yinTravelGalleriaSupportingRemembrance2025}. 

However, these prior VR works specifically \emph{design} for reflection --- they incorporate reflective guidance, artifacts, and scaffolding on top of an existing baseline. In contrast, embodied journaling emphasises the affordances of the isolated human body. Instead of designing \emph{in} VR, we use VR to strip away environmental surroundings and isolate embodiment --- with just one's body to express and reflect, what behaviours and interpretations emerge naturally? Through user studies, we explore the potential and drawbacks of embodied journaling in VR. We investigate how embodied journaling affects participants physically and emotionally, how it mediates the vividness of a lived experience, how it encourages (or discourages) reflective processes, and how it differs from written journaling. Our research was guided by two research questions:

\begin{itemize}
    \item \textbf{RQ1:} What behaviours emerge in embodied journaling in VR, and how do people reflect upon and interpret such behaviours?
    \item \textbf{RQ2:} How does embodied journaling in VR differ from written journaling in terms of affect, recollection, and reflection?
\end{itemize}

To exploratively evaluate the effects of embodied journaling, we conducted a within-subject comparative user study with $N=20$ participants consisting of two sessions. In the first session, participants engaged in the journaling process under both written and embodied conditions. Participants journaled and reflected upon an emotionally-loaded negative experience, such as a conflict with a partner, as these are situations traditionally served by written journaling \cite{dizon2020ect}. In the second session, they revisited their previously created artifacts (through re-reading and watching a recording).

We find that, in written journaling, participants captured \emph{precise descriptions} of the entire experience. Participants lingered, re-read, and picked out specific words to process the negative event. In embodied journaling, participants created \emph{abstract emotional representations} of the experience. They developed a vocabulary of motion, which was later interpreted back to their feelings during the negative event. While written journaling interleaves expression, processing, and self-presentation, embodied journaling focuses on raw expression first and meaning-unravelling afterwards, shifting towards \emph{reflection-on-action} from \emph{reflection-in-action}. We discuss how decentering reflective technologies from written expression can create new forms, temporalities, and outcomes of reflection. 

Altogether, this paper contributes: (i) a concept and functional prototype of embodied journaling in VR, (ii) a comparative evaluation against written journaling to show how embodied journaling supports emotional processing, reflection, and introspection, (iii) an understanding of how linguistic vs. non-linguistic modalities affect emotional expression and reflection, and (iv) suggestions on how embodied journaling can extend existing reflective practices. 

\section{Related Works}

\subsection{The Importance of Reflective Technology} 

Reflection --- thinking about action and understanding one's role within existence --- shifts introspective exploration into outward transformation \cite{atkins1993reflection}, allowing people to make sense of and adapt to the world and their experiences. Reflection is associated with many positive outcomes~\cite{stein2014}: facilitating problem-solving and creative processes \cite{krossSelfReflectionWorkWhy2023}, fostering self-insight~\cite{baumer2014reflection}, supporting life changes~\cite{Staudinger_2001}, and improving health, well-being, and personal growth~\cite{bryant2005using, lyubomirsky2005pursuing, slovak2017ref-practicum, Mols2020EverydayDott}. 

Most widespread is the foundational work on reflective practice by Schön~\cite{SchonThePractitioner, nanwani2021self-discovery}, distinguishing between reflection-in-action --- the spontaneous ability to “think on our feet” during interaction --- and reflection-on-action, which occurs retrospectively~\cite{SchonThePractitioner}. Bolton and Delderfield~\cite{bolton2018reflection} note that reflection-on-action enhances reflection-in-action. To understand what reflection entails, Baumer highlights three dimensions: breakdown (challenging existing concepts), inquiry (re-examining prior learnings), and transformation (reordering conceptual schema to enable change) \cite{baumerReflectiveInformaticsConceptual2015a}. Fleck and Fitzgerald offer a framework comprising levels of reflection to categorise the depths and potential outcomes of reflection, from merely describing events (Level 0), to reinforcing views (Level 1), shifting perspectives (Level 2), transforming thinking and behaviour (Level 3), and considering societal implications (Level 4)~\cite{fleckReflectingReflectionFraming2010}. 

Reflection can be a challenging activity and often does not occur automatically, instead needing to be encouraged externally~\cite{slovak2017ref-practicum}. Furthermore, engaging in reflective practice also bears the risk of getting stuck in worries and negative thought cycles, termed rumination \cite{krossSelfReflectionWorkWhy2023, harringtonInsightRuminationSelfReflection2010, marinRuminationSelfreflectionStress2017, eikeySelfreflectionIntroducingConcept2021}. As such, reflection as a nuanced design metric has been an enduring topic within HCI~\cite{sengersReflectiveDesign2005}. Reflection can be supported through providing conversations or prompts (e.g.~\cite{wagenerSelVReflectGuidedVR2023a}), reframing (e.g.~\cite{lukoffAncientContemplativePractice2020, wagenerMoodShaper2024, yinTravelGalleriaSupportingRemembrance2025, jiangPOVRDesigningImmersive2024}), ambiguity in design (e.g.~\cite{wagenerVeatherReflectEmployingWeather2023}), and leveraging temporality~\cite{hallnasSlowTechnologyDesigning2001a, odomExtendingTheorySlow2022a}. Bentvelzen et al. discuss four aspects of temporality --- past (revisiting prior experiences), memories (subjective reconstruction of experiences), future (taking artifacts into the future), and slowness (slowing down using technology) \cite{bentvelzenRevisitingReflectionHCI2022}. The concept of embodied reflection also introduces modality as a dimension of reflection. Dance practitioners, for example, can tune into their body and immerse themselves in their performance, transforming movements into attunement and learning \cite{schwenderCreatingFramingReflecting2025, buttingsrudBodiesSkilledPerformance2021}. 

Building on prior work on temporality and modality as a design dimension for reflection, our work explores how embodied journaling in VR captures the four aspects of temporality~\cite{bentvelzenRevisitingReflectionHCI2022} (past, memories, future, slowness), facilitates both reflection-in-action and reflection-on-action~\cite{SchonThePractitioner}, materialised embodied forms of introspection \cite{buttingsrudBodiesSkilledPerformance2021}, and supports different levels of reflection~\cite{fleckReflectingReflectionFraming2010}. 

\subsection{Journaling as a Reflective Practice} 

Journaling has long been recognised to support reflection \cite{hubbsPaperMirrorUnderstanding2005, utleyTherapeuticUseJournaling2011a, eppValueReflectiveJournaling2008, fritsonImpactJournalingStudents2008a}. Journaling entails a conversation with oneself \cite{cowanFacilitatingReflectiveJournalling2013a}, and this inner dialogue connects thoughts, feelings, and transformative actions \cite{hubbsPaperMirrorUnderstanding2005}, helping people process (negative) life events \cite{ullrichJournalingStressfulEvents2002c}. Journals reveal change over time, allowing people to conceptualise and understand the broader world \cite{sendallJournallingPublicHealth2013a}. Correia and Bleicher \cite{bleicherUsingSmallMoments2011} outlined how even journaling about small moments can potentially draw out important insights. Journaling offers opportunities to reinterpret past experiences into future learning \cite{elsdenItsJustMy2016}, inspiring a comparatively large body of HCI research in the domain \cite{karaturhanCombiningMomentaryRetrospective2022b, wangDesigningAIAugmentedJournaling2025a, bhattacharjeeActuallyCanCount2024a}. Much past research in HCI has looked to encourage this practice through incorporating journaling-like practices through digital mediums like phone apps, taking advantage of digital media affordances such as uploading photos, tagging, and filtering \cite{elsdenItsJustMy2016, karaturhanCombiningMomentaryRetrospective2022b, yin2025traveltales}. Increasingly, the design of journals has started to incorporate AI support, to help with processing and wellbeing \cite{kimDiaryMateUnderstandingUser2024b, Jung2024MyListener, wangDesigningAIAugmentedJournaling2025a}.

Often, the practice of journaling uses \emph{writing} as the medium of expression~\cite{cowanFacilitatingReflectiveJournalling2013a}, allowing people to convey their experiences, ground them in space and time, and form a tangible compendium of events \cite{bleicherUsingSmallMoments2011, cowanFacilitatingReflectiveJournalling2013a}. Pennebaker's work on expressive writing highlights the strength of putting negative experiences into words, which was shown to improve well-being \cite{slatcherHowLoveThee2006, pennebakerTheoriesTherapiesTaxpayers2004, gortnerBenefitsExpressiveWriting2006}. Yet, writing is not the only modality of journaling. Sawhney et al. \cite{sawhneyAudioJournalingSelfreflection2018a} explored audio input for language-based journaling, finding that audio provides more spontaneous expression that comes from understanding unfiltered intentions and subtle changes in tone \cite{sawhneyAudioJournalingSelfreflection2018a}. Arts-based and visual-based journaling employ abstract images to construct meaning and engagement \cite{redmondArtAudiencingVisual2022, lajevicTographyEthicsEmbodiment2008}. Abstract representations of experiences (e.g. in VR) can help in journaling beyond words, such as through environmental interaction and construction \cite{wagenerSelVReflectGuidedVR2023a, wagenerMoodShaper2024}, emotion and mood representations \cite{wagener2022MoodWorlds, yuCreativelySupportingMental2025}, and physical facial expressions \cite{jangDesignFieldTrial2024}. These works indicate that emotions and narrative emerge differently based on varying modalities of journal-like expression. 

Inspired by these findings and by somaesthetic design \cite{hookDesigningBodySomaesthetic2018}, we explore the medium of body (consisting of motion and speech) as a modality for journaling. Somaesthetics focuses on the body's role in sensory cognition, perception, and self-expression \cite{shustermanSomaestheticsDisciplinaryProposal1999, shustermanBodyArtsNeed2012}, highlighting the potential in people's felt stances~\cite{hookDesigningBodySomaesthetic2018}; somaesthetics has drawn interest in HCI as a way of designing with and for the body \cite{hookSomaestheticAppreciationDesign2016, lokeSomaticTurnHumancomputer2018}. We also drew inspiration from prior work indicating that motion can potentially represent experiences and support reflective processes \cite{habana2023embodied, wangItsMeVRbased2022, mauMentalMovementsHow2021}. For instance, the Self-Care Technology Process Model (SCTpm) emphasises the importance of motion when designing VR experiences for reflection~\cite{wagenerSelfCareTechnologyProcess2025}, which extends prior understandings of embodied reflection \cite{schwenderCreatingFramingReflecting2025, buttingsrudBodiesSkilledPerformance2021}. Combining both, we formalise what it means to ``journal'' using the human body. 

\subsection{The Potential of VR for Reflection} 

Although embodied journaling was a somaesthetic conceptualization, the medium of virtual reality (VR) offers several affordances that make it appealing in supporting the process. VR can immerse users in realistic interactive worlds with new perspectives \cite{chanEffectsVirtualReality2020, thamUnderstandingVirtualReality2018a}, providing a strong sense of presence \cite{schultzeEmbodimentPresenceVirtual2010b}. VR foregrounds the senses of self-location, agency, and body ownership, heightening the sense of embodiment \cite{guySenseEmbodimentVirtual2023, kilteniSenseEmbodimentVirtual2012c} as users inhabit a virtual body (an avatar) as a second self \cite{freemanMyBodyMy2020b}. With these affordances, users respond to stimuli naturally in VR, even while understanding that it is an illusion \cite{slater2018immersion}. Taking advantage of this potential, prior works have explored eliciting and mediating emotional states through invoked virtual scenarios \cite{liuPositiveAffectNatural2023, banosChangingInducedMoods2006, susindarFeelingRealEmotion2019, zulkarnainSelfAssessedExperienceEmotional2024, gallEmbodimentVirtualReality2021a}.

However, Jiang and Ahmadpour argue that VR's immersion can sometimes detract from opportunities for self-awareness and critical reflection \cite{jiangImmersionDesigningReflection2022a}. They outlined how reflective VR systems can be grounded in specific design strategies, including viewpoint manipulation and immersive estrangement \cite{jiangImmersionDesigningReflection2022a, jiangPOVRDesigningImmersive2024}. Many reflective VR systems have adapted variants of such strategies. For example, meditation and mindfulness-related tools take advantage of VR's sensory escapism and immersion \cite{feinbergZenVRDesignEvaluation2022d, pengMetaFlowExperience2023, wangReducingStressAnxiety2024, kumarVRZMExploringPsychophysiological2024, yanMindBodyTaoRelaxRelieving2024}; other systems may elicit reflection through encouraging prompts~\cite{wagenerSelVReflectGuidedVR2023a}, conversations with friends and family~\cite{stefanidi2024TeenWorlds, wagenerTogetherReflectSupportingEmotional2025}, memory curation \cite{yinTravelGalleriaSupportingRemembrance2025, bahngDesigningImmersiveStories2023a}, metaphoric reframing~\cite{grieger2021TrashIt, raschMindMansionExploring2024, wagenerMoodShaper2024}, alternative perspective-taking \cite{bahngReflexiveVRStorytelling2020, jiangPOVRDesigningImmersive2024}, or ambiguous and qualitative visualisations~\cite{wagenerVeatherReflectEmployingWeather2023}.

Our work deconstructs typical guided reflection in VR, providing only minimal guidance and feedback. Instead, we focus specifically on what Jiang and Ahmadpour deem \emph{immersive estrangement} --- users are siloed and defamiliarised from the real world, and embody a nondescript avatar in a solitary world \cite{jiangImmersionDesigningReflection2022a}. Although people can observe themselves, as in prior works \cite{vuarnessonVirtualRealityInward2024, zhouVirtualRealityReflection2021}, their task mostly remains abstract and subjective (similar to Roo et al.'s Inner Garden \cite{rooInnerGardenConnecting2017}). Rather than actively designing for reflection, we are interested in how reflective processes emerge from people's embodiment. We explore how an isolated embodiment of the expressive human body \cite{hookDesigningBodySomaesthetic2018, hookSomaestheticAppreciationDesign2016, habana2023embodied} to represent a lived experience might mediate people's reflections and meaning-making experiences. 

\begin{figure*}[h]
  \centering
  \includegraphics[width=1\linewidth]{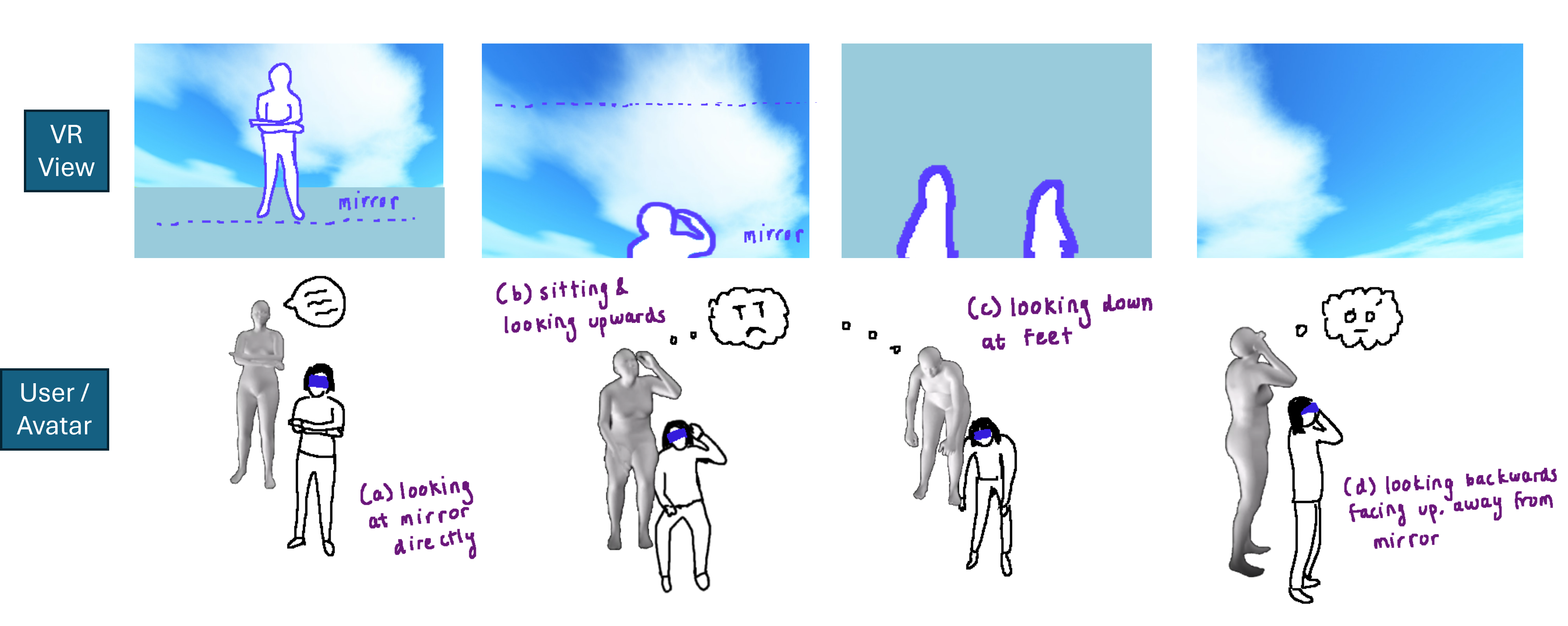}
  \caption{This hypothetical storyboard was inspired by a typical journey of reflection during embodied journaling as most commonly expressed by participants. It depicts typical body poses chosen in different stages of reflection during the process of embodied journaling. Each body pose depicted here was chosen by a participant of our study. The figure shows both the first-person view in VR (having a mirror) and the body poses as seen from outside. In (a), the user adopts an arms-crossed pose while talking about a difficult experience. In (b), the user sits and looks up, pondering the negative experience while looking around the virtual environment. In (c), the user looks down at their feet, and the participant reported feeling ``down'' at that moment. In (d), the user walks around, turns around, and looks up, gaining some clarity after thinking.}
  \Description{This image shows a comic-like storyboard of embodied journaling for the processes of creation and review. Each panel has both the VR view and a view of the user. In (a), the user adopts a neutral arms-crossed pose looking at the virtual mirror. In (b), the user sits and looks up, which is mirrored by the avatar. In looking up, in VR they see the top of the mirror and their head instead of the full body. In (c), the user looks down at their feet, and in VR they only see the avatar's leg and feet. In (d), the user is looking away from the mirror towards the sky. In VR, they no longer see their avatar, instead seeing the sky solely. }

  \label{fig:storyboard}
\end{figure*} 

\section{Designing Embodied Journaling}
The idea of journaling using the body, i.e., using embodied physical motion to \emph{represent}, \emph{process}, and \emph{review} experiences, is the novel process explored in this paper. We highlight that our focus is not on prescribing a specific workflow, but on understanding what practices and expectations emerge when people `write' with their body itself, similar to filling a page with words. 
In this section, we explain this idea through metaphors with more familiar written journaling and outline the development of our system that collects and replays embodied journal entries. 

\subsection{Leveraging VR for Embodied Journaling}

\begin{figure*}[h]
  \centering
  \includegraphics[width=0.95\linewidth]{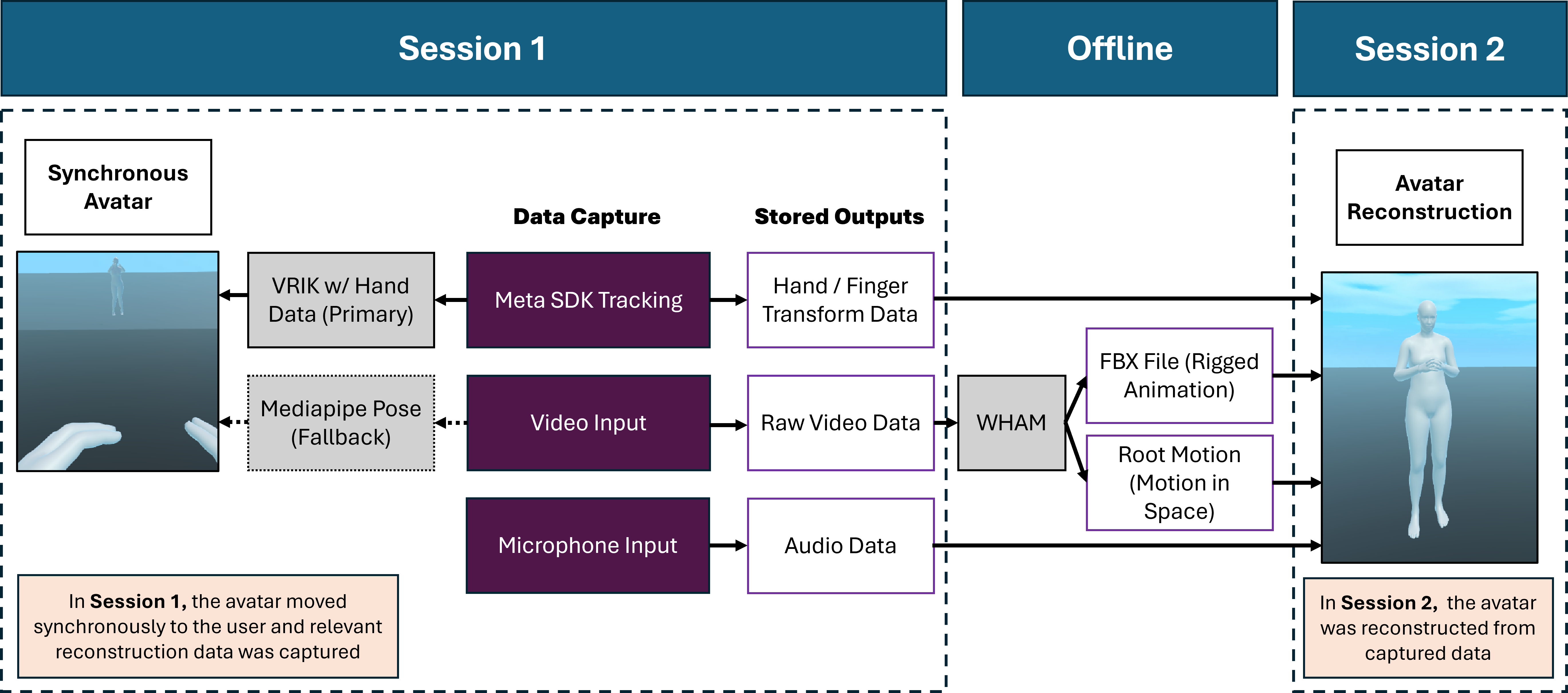}
  \caption{The full pose estimation pipeline for our embodied journaling system. In Session 1, we use existing tracking tools such as SDK tracking, along with synchronous algorithms such as VRIK and a Mediapipe fallback when the hands are out of view, to render the avatar pose in real-time. During session 1, we also collect movement data, processed offline using WHAM to generate more accurate animations. In Session 2, we use these reconstructed avatars to accurately replicate the initial journal.}
  \Description{This image shows the pose estimation pipeline for our embodied journaling system. In session 1, the system captures data from Meta SDK tracking, video input, and microphone input. The SDK tracking is used for the real-time avatar movement with the VRIK algorithm. However, the video input with the Mediapipe Pose tracking is used as a fallback when tracking is lost. From this data, stored outputs in the form of hand and finger transform data, raw video data, and audio data are extracted. Between the two sessions, the researcher processes the raw video data into an FBX file with the rigged animation of the participant and their root motion within 3D space. Using all of these data points allows the system to reconstruct the avatar accurately in the second session. }

  \label{fig:pose}
\end{figure*}

We implement embodied journaling using VR, due to VR's essential affordances in:

\begin{itemize}
    \item \textbf{Embodiment}: VR exercises a high degree of embodiment \cite{kilteniSenseEmbodimentVirtual2012c} for users to inhabit and explore using a virtual body. In our system, users control a full-body avatar, and self-awareness is made obvious through the use of a virtual mirror. We used a nondescript, basic model without facial expression (SMPL \cite{SMPL:2015}) rather than a realistic mirroring of the user because we wanted to consistently emphasise embodied action as expression rather than identity, appearance, or communication cues that could affect emotion and perception \cite{zhouVirtualRealityReflection2021}. Although a similar setup could be employed in non-VR, the abstraction of a person's individual appearance into a nondescript model allows us to remove appearance-based effects and biases \cite{freeman2021body, negrinAppearanceIdentity2008, naumannPersonalityJudgmentsBased2009}, which would be inevitable if a person were to view themselves in the real world. 
    \item \textbf{Environmental Isolation}: VR silos the users from the real world, immersing them in a virtual environment. In the VR setup, users are transported into and navigate a minimalistic, solitary space (further explained in Section \ref{sec:space}) free from distractions and external factors. We believed this would centre their focus on their experience and feelings they were journaling. On the other hand, a non-VR set-up could be liable to real-world environmental distractions, which could compromise memory \cite{varao-sousaLabWildHow2018}. 
\end{itemize}

In addition to making the embodied journals in VR, we also considered how people might review them. Similarly to how users might re-read their written journals as additional processing and reflection, users can also watch how they moved and what they said from a third-person perspective in VR, i.e., watching their avatar from embodied journaling. For the review phase, we leveraged the following affordance of VR:

\begin{itemize}
    \item \textbf{Depth and Movement Trajectories}: VR allows for depth understanding, emphasizing the motion of the avatar as a 3D being within 3D space. In contrast, flat 2D video of a person's movements lacks depth information, creating challenges in understanding the intricacies of motion \cite{mohrRetargetingVideoTutorials2017, gerigMissingDepthCues2018}. 
\end{itemize}

We aimed to keep our system simple, lightweight, and accessible with off-the-shelf hardware, requiring a VR headset and a camera (or webcam or phone), without the need for expensive trackers or added sensors.

\subsection{Implementation of Embodied Journaling}

We developed our embodied journaling system using Unity version 2022.3.30f1, interfacing with a Meta Quest 3 using the Meta XR SDK\footnote{\url{https://developers.meta.com/horizon/develop/unity}} and running the project via Meta Quest Link. The system was developed on a Windows computer with an NVIDIA GeForce RTX 3080 graphics card, and the experiments were run on a Windows laptop with an NVIDIA GeForce RTX 3060 Laptop graphics card. When the project starts up, the user is transported into an initial scene, which differs based on whether the user is creating a journal or reviewing one: 
\begin{itemize}
    \item \textbf{Creation Scene}: The creation scene is focused on experiencing and exploring embodied journaling, but also helps in collecting reconstruction data as well. The user sees a large virtual mirror occupying the virtual space, with an embodied avatar mirroring their motion. As users make their journal, we collect and record data --- the person's audio, their granular finger motion, and a video of their body. Audio and finger motion are directly retrieved from the Unity scene itself (via the microphone and hand tracking), and a video of the person's body is retrieved through a webcam; in particular, we used a Samsung Galaxy S25 connected to the computer as an external webcam using Iriun Webcam\footnote{\url{https://iriun.com/}}. Figure \ref{fig:storyboard} shows a stylised drawing of journal creation. 
    \item \textbf{Review Scene}: The review scene is focused on replaying the recorded journal. Here, the user sees a virtual avatar that mimics their motion and speech from their initial recording, with an option to play and pause. In the study, while participants watched this recorded avatar, they started two metres away (in-world) and level in height while standing. However, users could also move closer and further to the avatar by moving around; furthermore, we highlight that all study participants chose to sit out of comfort during review, thus often viewing the avatar from slightly below. 
\end{itemize}

\subsubsection{Designing the Virtual Space and the Mirror}
\label{sec:space}

Constructing the virtual space for embodied journaling was vital. We wanted the space to be minimalistic to offset distractions \cite{varao-sousaLabWildHow2018}. After brainstorming within the research team, we settled on a minimalist world with a sky background and a plane for the avatar to stand on (see Figures \ref{fig:teaser} and \ref{fig:pose}). This solitary space would serve as the metaphorical canvas for written journaling, which also typically takes place on a blank page. In written journaling, it is people's words that fill the empty page with experience and meaning; in VR, we envisioned that people's actions might fill the physical canvas similarly. 

During the embodied journaling session (i.e., journal creation), we also decided to incorporate a mirror. The mirror extends beyond room-size and stays in place, statically anchored in the environment --- users can choose to observe themselves, their gestures, their movement, and embodied actions as they move around and speak, or they can look away. Metaphorically, the mirror serves as a reflection of one's actions, just as words on a page or in a text document might reflect one's words. The choice of having a mirror comes with practical advantages too, as it could provide visual feedback to participants while offering a way to focus on self-movement \cite{marquardtSuperMirrorKinect2012, krugMirrorMeExploring2025, fuAmMirrorDweller2023, andersonYouMoveEnhancingMovement2013}. During our testing, we found that the mirror helped us preview and focus on our motion; without it, there was a distinct lack of visual feedback. Still, mirrors also pose potential drawbacks in our journaling context that can be explored in future designs, as the increased focus on movement can also accompany increased awareness and self-consciousness \cite{fuAmMirrorDweller2023}.

\subsubsection{Pose Estimation}

Given the focus on embodiment in embodied journaling, the virtual model needed to accurately mirror the user's motion. Users were represented using the blank SMPL model (which could be female or male, depending on user-selected preference) \cite{SMPL:2015}. The SMPL model, as an FBX file, was linked to the motion of the avatar in Unity using FinalIK\footnote{\url{https://assetstore.unity.com/packages/tools/animation/final-ik-14290}}, an inverse kinematic solution for Unity. In particular, we used the VRIK solver to configure avatar body tracking given the user's head and hand positions and rotations during journal creation. This means that the solver used the head and hand data (which were always spatially accurate and could be obtained from the Meta XR SDK\footnote{\url{https://developers.meta.com/horizon/develop/unity}}), to estimate other limbs (such as torso, legs and feet).

Relying solely on this VRIK solver does not capture minute finger motion, even though we already had a real-time, synchronous hand tracker --- Meta XR's innate hand tracking. To achieve granular finger motion with full-body tracking, we removed the original model's hands and ``stitched on'' Meta XR's hands. However, this only functions optimally when the hands are in the field-of-view (FoV) of the Quest. When the hands were outside the FoV, we instead fell back to pose estimation via Mediapipe pose estimation\footnote{\url{https://ai.google.dev/edge/mediapipe/solutions/vision/pose_landmarker}, via \url{https://github.com/ganeshsar/UnityPythonMediaPipeAvatar}} (using the concurrent video recording) to understand where to place the hands, although the finger motion was lost during these moments. 

Synchronous mapping of the user to the avatar was not perfect during journal creation. Inverse kinematics for pose is imperfect, the hands are not always fully attached to the model, and the hand fallback loses granular information when outside of the Quest's FoV. Our system provides fallback mechanisms for graceful degradation as data sources become unavailable. We heuristically fine-tuned the thresholds for each fallback until our system subjectively performed well synchronously.

For the asynchronous replay of the embodied journal taking place afterwards, we took advantage of the offline time to run a high-quality pose reconstruction algorithm. We used the WHAM algorithm \cite{shin2023wham} to reconstruct a user's three-dimensional motion and pose through their recorded full-body video\footnote{\url{https://github.com/yohanshin/WHAM}}, which we then converted to an FBX file. This process generally took several hours on our RTX 3080 machine, depending on the length of the video. As WHAM is not granular to the finger level, we again replaced the model's hands with Meta XR's hands, having finger motion captured during the initial recording. Overall, we found this process produced an accurate representation of their embodied journal, which incorporated not just pose, but detailed finger motion as well. The body pose estimation pipeline is illustrated in Figure \ref{fig:pose}.

\section{User Evaluation}
We conducted a comparative exploratory user study (approved by our institute's ethical review board) to evaluate embodied journaling. Through within-subject comparison with traditional written journaling, we uncovered how reflective practice emerges in similar and different ways through embodied VR. Despite the limitations of within-subject comparison, such as complicating ordering and incurring carryover effects \cite{hoCausalInferenceCounterbalanced2025}, this methodological design allowed us to elicit exploration and provide a baseline for participants to judge against. Unlike a between-subject study, where one condition would be invisible and qualitative findings would be siloed to a single condition, participants were instead able to compare and contrast how they felt across conditions.

We primarily focused on the journaling of \emph{resolved, negative} experiences. Journaling on negative events aligns strongly with prior applications of the practice in literature and focuses on its ability to induce positive effects. We opted for resolved experiences --- experiences in which participants had already worked through and reflected upon --- because we wanted to be cognizant of people's well-being and minimise potential emotional harm. 

The study comprised two sessions. The first session focused on creating the reflective artifacts regarding the negative experience, and the second session focused on reviewing them (see Figure \ref{fig:study}; full study protocol is in the supplementary material). The sessions were scheduled 6--9 days apart, which we judged to be sufficient to allow temporal distance between creation and review. Each session took approximately 1 hour, and all studies took place at the primary researcher's institute. Before each session, the researcher introduced the research, described data collection and usage, and asked participants to fill out a consent form. Participants were compensated at a rate of \$16 CAD per hour. 

\begin{table*}[h]
  \caption{Study Participant Demographics. VR Experience and Journaling were scored on 4-point Likert scales ([\emph{No Experience / Beginner / Intermediate / Advanced}] for VR; [\emph{No Experience / Beginner / Intermediate / Frequently}] for Journaling)}
  \label{tab:participants} 

    \begin{tabular}{|c|c|c|c|c|}
    \hline
    Participant ID & Age & Gender & VR Experience & Journaling Experience \\
    \hline
    \rowcol 1 & 28 & Man & Advanced & Beginner\\ 
    \rowwhi 2 & 23 & Man & Intermediate & Frequently\\ 
    \rowcol 3 & 24 & Man & Beginner & Beginner\\ 
    \rowwhi 4 & 22 & Woman & No Experience & Beginner\\ 
    \rowcol 5 & 31 & Woman & Beginner & Frequently\\ 
    \rowwhi 6 & 27 & Man & Advanced & Intermediate\\ 
    \rowcol 7 & 28 & Woman & No Experience & Intermediate\\ 
    \rowwhi 8 & 23 & Man & Advanced & Beginner\\ 
    \rowcol 9 & 27 & Woman & Beginner & Beginner\\ 
    \rowwhi 10 & 25 & Man & Advanced & No Experience\\ 
    \rowcol 11 & 21 & Man & Beginner & Beginner\\ 
    \rowwhi 12 & 20 & Woman & No Experience & Frequently\\ 
    \rowcol 13 & 18 & Woman & Beginner & Beginner\\ 
    \rowwhi 14 & 19 & Man & Beginner & Intermediate\\ 
    \rowcol 15 & 35 & Man & Beginner & Beginner\\
    \rowwhi 16 & 28 & Man & Intermediate & Beginner\\ 
    \rowcol 17 & 24 & Woman & Beginner & Intermediate\\ 
    \rowwhi 18 & 19 & Woman & Beginner & Beginner\\
    \rowcol 19 & 28 & Woman & Advanced & No Experience\\ 
    \rowwhi 20 & 19 & Woman & No Experience & Beginner\\ 
    \hline
\end{tabular}
  \Description{Demographic details of study participants, with participant ID, age, gender, VR experience, and journaling experience}
\end{table*}

\subsection{Data Collection}
We collected quantitative data regarding affective, recollective, and reflective dimensions through standardised questionnaires, and qualitative data through interviews at the end of each session.  

\subsubsection{Measures}
In the first session, we used the \emph{Positive and Negative Affect Schedule (PANAS)} \cite{watsonDevelopmentValidationBrief1988} to assess participants' affective qualities. We used this to evaluate how their emotions changed before and after creation in embodied and written journaling. In the second session, we used the shortened \emph{Memory Experiences Questionnaire (MEQ-SF)} \cite{luchettiMeasuringPhenomenologyAutobiographical2016} to assess the ability of each condition's journal entry to support recollection of the participants' negative experience. This scale measured memory-based qualities such as vividness and temporal perspective.

In both sessions, we used the \emph{Reflection, Rumination, and Thought in Technology Scale (R2T2)} \cite{loerakkerTechnologyWhichMakes2024b} to understand how embodied and written journaling supported reflective qualities in both creation and review across embodied and written journaling. In this first session, the scale term \emph{``this technology''} referred to the process of creating the entry; in the second session, the term referred to the process of reviewing the entry. 

\subsubsection{Interview Protocol}
At the end of the first session, we conducted a semi-structured interview, asking participants about how they felt in each condition and what they expressed through each modality (including if and why they spoke). Interviews took between 4--19 minutes, with an average of 12 minutes. At the end of the second session, we again conducted a semi-structured exit interview with participants, asking about how they felt reviewing each condition, how they interpreted their artifacts, and to what degree they felt connected to their previous self. The full interview guideline for both sessions can be found in the supplementary material. These second interviews took between 13--33 minutes, with an average of 20 minutes. 

\subsection{Participant Sample}

Participants were recruited through a mix of convenience sampling and a paid study listing made on our institute's study board. In total, $N=20$ participants (10 females, 10 males, with an average age of 24.5, $min: 18$, $max: 35$; see Table \ref{tab:participants}) took part in the study. We queried participant experience with VR experience and journaling frequency, and administered the \emph{Self-Reflection and Insight Scale (SRIS)} \cite{grantSELFREFLECTIONINSIGHTSCALE2002a} to get insights about the reflective capacities of our participant sample. We found that our sample scored an average of 90.1 on the SRIS (out of 120), ranging from 75 to 105 with a standard deviation of 8.14, which we interpreted to mean a high degree of self-reflection in our sample. Before the study, participants were asked to think of a resolved negative experience to journal about, as well as to confirm again that they felt mentally ready to recount this negative experience. 

\subsection{Study Procedure}

\begin{figure*}[h]
  \centering
  \includegraphics[width=1\linewidth]{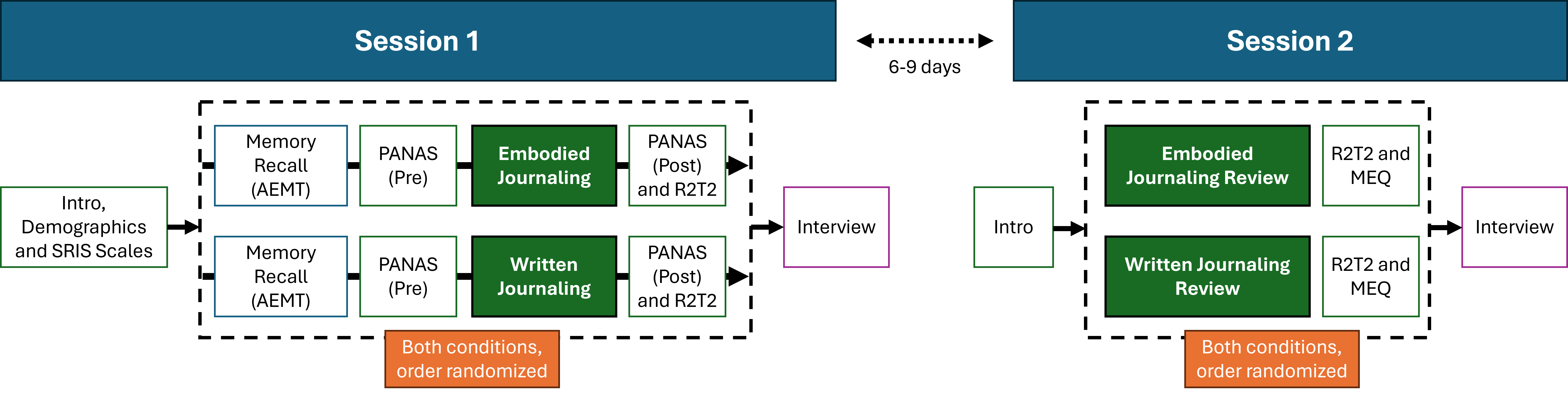}
  \caption{A visualization of the study procedure across the two sessions.}
  \Description{In this image, we see a flowchart of the study protocol across the two sessions. On the left, we start with Session 1. It begins with the researcher's intro, demographics survey, and SRIS, before splitting into the order-randomised conditions. Each condition involves a memory recall task (AEMT), the pre-condition PANAS, the journaling condition (embodied or written), the post-condition PANAS, and the R2T2 scale. After both conditions, Session 1 ends with an interview. Session 2 begins 6--9 days after Session 1. After an introduction, participants are split into the order-randomised conditions of reviewing their embodied journal or their written journal. After each, they are asked to fill out the R2T2 scale and the MEQ-SF scale. At the end, another interview is conducted. }
  \label{fig:study}
\end{figure*}

\subsubsection{Session 1 --- Creating the Journal}

The first session focused on creating the journal entry, looking specifically at \emph{how embodied expression emerges from user interpretation}, and \emph{how it feels different} from written practice. Participants provided informed consent and filled out the demographic data survey and the \emph{SRIS}. 
    
In our within-subject study, participants experienced two conditions with a similar setup. First, participants recounted a negative, emotional experience to the researcher through a prompt adapted from the \emph{Autobiographical Emotional Memory Tasks (AEMT)} to bring them back to the experience \cite{aldama2019use}, after which they filled out the PANAS. Participants were then asked to journal on that negative experience either through written journaling or embodied journaling in VR --- the same experience was used in both conditions. The order in which they performed each type of journaling was counterbalanced across participants. For each condition:

\begin{itemize}
    \item \textbf{Written Journaling}: Participants journaled using written text, typing into a textbox (built on Qualtrics) on a computer screen using a mouse and keyboard. After writing, they were asked to click a submit button, and the written text was directly recorded. 
    \item \textbf{Embodied Journaling}: Participants journaled in VR using our system, exploring the space using an embodied avatar.
\end{itemize}

The task instructions for each condition involved asking the participant to journal while the researchers generally remained silent. Given the novelty of the embodied experience, we expected that participants might not always know how to start (which proved true --- several participants initially expressed some initial uncertainty). Thus, during the task instruction, we offered examples of actions that could be done during embodied journaling, e.g., gesturing, walking, being still, being silent, or speaking. 

In both conditions, participants were told to take whatever time they needed to complete the task; however, we would verbally give a gentle nudge around 10 minutes to keep the study on time. After each condition, participants filled out another \emph{PANAS} scale and the \emph{R2T2} scale, and between the conditions, we asked participants to take a 2-minute break to mentally reset. We conducted a semi-structured exit interview after both conditions had been completed. 

\subsubsection{Session 2 --- Reviewing the Journal} 

Before the second session, we processed the video to generate the participant's VR replay and transferred their written journal into an electronic document. Then, the second session focused on looking back on these two reflective artifacts, looking at \emph{how they differed in terms of affective, recollective, and reflective qualities}. For each condition, participants revisited and reflected on their previous entry. The order in which they reviewed was counterbalanced across participants. For the:

\begin{itemize}
    \item \textbf{Written Condition}: We asked participants to read and process their written journal.
    \item \textbf{VR Condition}: We asked participants to watch through their embodied avatar in virtual space while wearing the VR headset, considering their motion and auditory capture. 
\end{itemize}

In both conditions, participants were given the aligned instructions of exploring their thought process, the content of the entry, and how they felt. Participants were told they could take as much time as they needed, and the researchers were silent during this task. After each condition, participants were asked to fill out the shortened \emph{Memory Experiences Questionnaire (MEQ-SF)} and the \emph{R2T2} scale; between the conditions, we asked participants to take a 2-minute break to mentally reset. After both conditions, we conducted a semi-structured exit interview.

\subsection{Study Set-Up}
During the study, participants could freely manoeuvre in a 2.5x2.5 metre area. We turned off the Guardian system to maintain immersion and planned to have the researcher call out if participants wandered too close to the physical boundary. Although the space was cleared out, a small number of participants mentioned that they wanted to use a chair at the boundary of the space during journaling. This chair was not rendered in VR --- participants relied on both memory and the headset's nose gap for spatial recall.

\subsection{Data Analysis}
We follow a mixed-methods approach, with qualitative results at the forefront. All audio recordings of the anonymised interviews were transcribed verbatim and imported into Airtable\footnote{\url{https://airtable.com/}}. We began with informal familiarization of the data itself, discussing initial trends and interesting findings that we observed. We then moved to a more structured thematic analysis \cite{braun2021TA}. To align the collaborative process, two authors both independently coded both sessions' interviews for the same three participants through initial open coding~\cite{braun2006thematic-analysis}.

Through iterative discussion of these initial interviews, a preliminary codebook and coding tree were established. The remaining transcripts were coded individually by one author based on this initial codebook, although the codes were still subject to evolution and shifting as both coding and discussion continued. To synthesise findings, we followed Blandford et al.'s pragmatic approach to thematic analysis~\cite{blandford2016qualitative}. In particular, two authors performed a collaborative affinity mapping exercise on Miro\footnote{\url{https://miro.com/}} to identify themes through thematic analysis. We decided to intersect the data from both sessions, as findings from each session were inevitably tangled with the context of the other. By grouping the codes into broader hierarchical categories, we were able to develop patterns of findings, which became our themes. 

Quantitatively, the data was analysed through statistical tests. Based on visual inspection of our data and the Shapiro–Wilk statistic between matched groups, we could not assume normally distributed data. Thus, we chose to use non-parametric statistical tests. Wilcoxon tests were used to compare across paired groups of measures, shown in Table \ref{tab:stats}. One-way ANOVAs of Aligned Rank Transformation (ART) \cite{wobbrock2011aligned} were used to compare across both measures and sessions for the R2T2 data. Our full data is presented in the supplementary material. 

\section{Qualitative Findings}

\subsection{Expression in Journaling}

As a gateway into the rest of the findings, it is important to understand what the \emph{shape} of embodied journaling looked like, especially for such a novel practice. To start to address \textbf{RQ1}, our qualitative analysis first looked at people's embodied actions and their interpretation of these actions.

\subsubsection{Effects of Language-Based Expression}
\label{sec:effectslanguage}
This section examines participants’ experiences with writing down their thoughts versus speaking them out loud in the embodied journaling condition, as both are language-based forms of communicative expression. We wanted to understand the linguistic and emotional differences observed between these two modalities.  
We will preface by stating that not every person chose to speak in the study; in fact, the majority of the 20 participants refrained from talking ($N = 13$). This was somewhat surprising to us, as we initially thought that people might want to translate their written thoughts directly to speech --- P17, who did speak, mentioned that \emph{``I decided to talk, which is the next closest thing to journaling and writing down my thoughts''}.

Thus, we asked participants who did not speak, why? For several participants, speaking often came into tension with thinking, and they were primarily focused on the latter during the study. For instance, P19 mentioned that \emph{``being quiet makes me more conscious, and I can better focus on myself and my feelings compared to [talking]''} and P6 (who was silent for most of the time) mentioned that \emph{``the throughput of thinking is much higher than talking''}. However, beyond this more grounded reason to refrain from speaking, participants also mentioned feelings of general awkwardness:

\begin{quote}
    \emph{``It was just awkward for me to speak... I feel comfortable writing to myself, but I've never spoken out loud to myself.''} - P9 
\end{quote}

\begin{quote}
    \emph{``I don't really talk about my things out loud to myself or others. I would usually just think about more words in my head.''} - P13 
\end{quote}

This also stemmed from uneasy feelings of non-privacy that speaking elicited, which made some participants feel vulnerable. 

Participants who did speak were asked to compare the contents of their speech and their writing during the interview, both being language-based forms of communication. The primary difference that participants brought up was that speaking was more raw, whereas writing was more filtered; participants pondered and polished the words that they wrote:

\begin{quote}
    \emph{``I felt like the VR journaling was a bit more stream-of-consciousness. It was a little more unrefined.... [In the writing] I sat there and thought about the words before typing them.''} - P7
\end{quote}

Despite subtle differences in how the experience was expressed in language, participants' perceptions of their speech differed from their writing primarily in paralinguistic cues rather than content.

\begin{quote}
    \emph{``You can hear changes in inflection, changes in the speed in which you're talking. You can hear when you make mistakes and correct yourself.''} - P3
\end{quote}

\begin{quote}
    \emph{``The tone of my voice tells a lot when I'm listening back on it. I feel I was more monotone, and it was more of a defeated tone.''} - P4
\end{quote}

However, even for participants who did not speak, there were still sometimes emotional auditory paralinguistic cues that writing would not capture --- for example, P8 brought up how they took a \emph{``deep breath''} when standing up, P16 brought up how they can hear \emph{``many sighs''}. 

\subsubsection{Effects of Motion-Based Expression}
\label{sec:effectsmotion}

\begin{figure*}[h]
  \centering
  \includegraphics[width=0.9\linewidth]{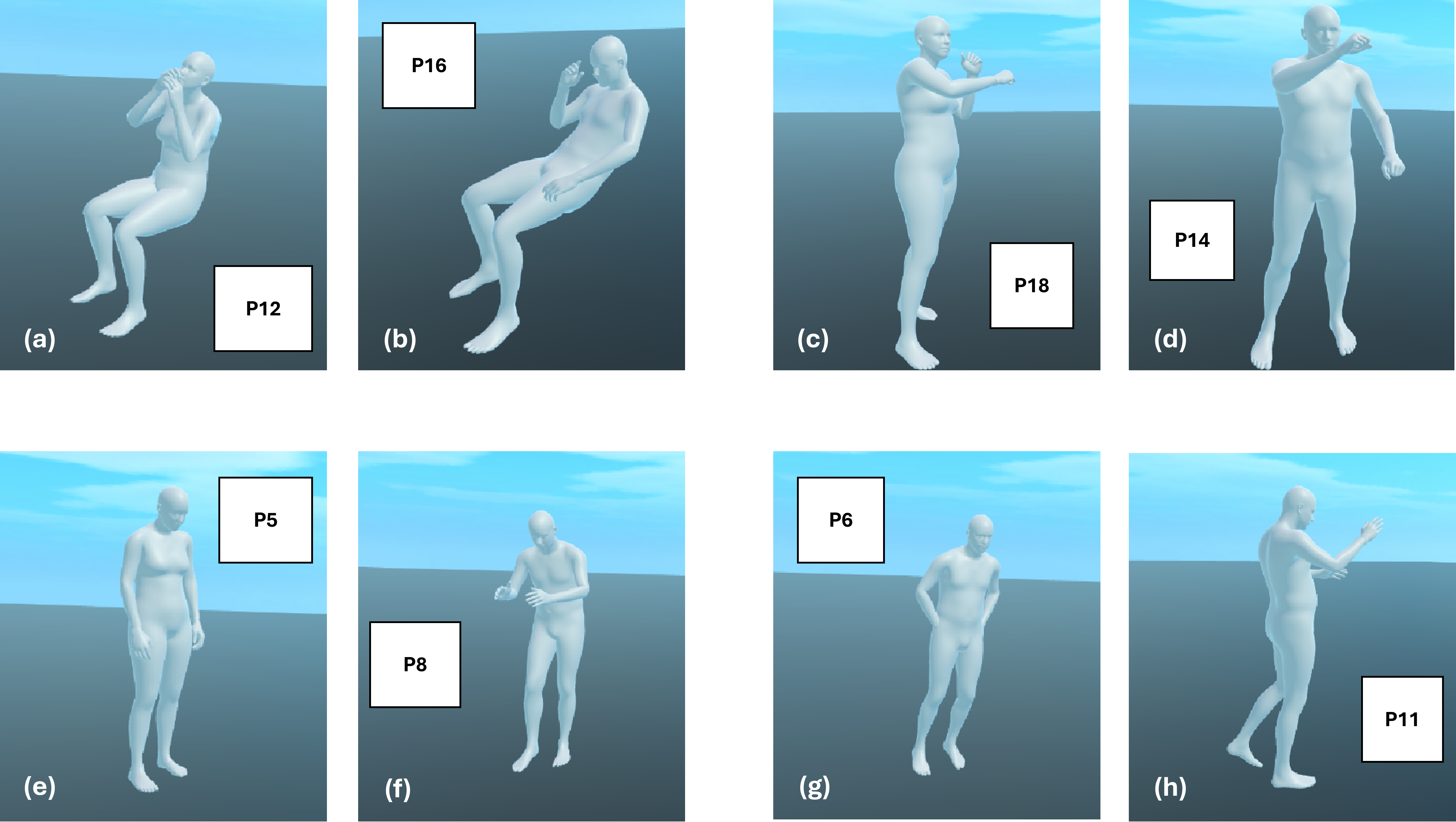}
  \caption{Still images taken of participants' embodied journals that demonstrate some shared patterns of motion. In (a) and (b), participants sat down; in (c) and (d), participants made punching gestures; in (e) and (f), participants gazed downwards and closed off their posture; in (g) and (h), participants paced around the room. See our video figure for more examples. }
  \Description{Here, we see 8 subimages drawn from participants' embodied journals, each involving a nondescript avatar doing some form of bodily expression. In (a), the avatar is sitting, with their hands touching their face and looking up. In (b), the avatar is sitting, with one hand on their face and looking down. In (c) and (d), the avatars are punching the air. In (e) and (f), the avatars are slouched and withdrawn, facing down. In (g) and (h), the avatars are in the process of pacing around. }
  \label{fig:actions}
\end{figure*}

We rendered and replayed each participant's body motion --- including moments of stillness --- with examples of various captured gestures displayed in Figure \ref{fig:actions}. 
Almost all participants remarked on how determining how to move was initially more challenging than what to write, especially as the latter has stronger established norms. However, participants developed their own rhythms over time, and we observed a wide range of movement behaviours. In terms of macroscopic patterns --- the large, goal-based actions that participants were generally aware of --- spanned: 

\begin{itemize}
    \item{\textbf{Contemplative Motion}}. Their thoughts about the experience naturally guided their motion. Participants described being still --- \emph{``when I was still, I was just thinking about my memory''} (P18), pacing around --- \emph{``I pace to get my memory as well as those words out more''} (P11), and sitting down --- \emph{``to calm myself down and collect my thoughts''} (P12).
    \item{\textbf{Re-enactment}}. They used motion to revisit the experience through acting out the events. For instance, they were --- \emph{``describing [through movement] how we get up on the mountain''} (P1) and \emph{``trying to explain the summary with my gestures''} (P9). Sometimes re-enactment was used to augment what really happened to express \emph{what-ifs?} P14, who re-enacted meeting the people largely responsible for their negative experience, used punching gestures to express having \emph{``punched them hypothetically''} (P14).
    \item{\textbf{Interpreted Expression}}. Participants used motion to \emph{consciously} and \emph{purposefully} express their feelings, for instance, through dance or gestures. One participant interpreted their feelings through a \emph{``contemporary form of dance, where we show expressions through body expansion and contraction''} (P5), and P3 mentioned using \emph{``hands to draw a graph of changes in emotional balance and arousal over time''} (P3). 
\end{itemize}

Within these macroscopic patterns, smaller microscopic motions occurred, such as changes in posture, head motion, or gestures. Sometimes participants were not even initially aware of these movements, and only realised them when reviewing their motions in the second session. Participants later interpreted these subtle cues:

\begin{quote}
    \emph{``I didn't really notice... I looked down a lot...  I think when I'm looking down, I'm trying to think more about the situation.''} (P13)
\end{quote}

\begin{quote}
    \emph{``It's very much just hunched, and I can see that my hands are here. They're closed off. They're protecting my insides.''} (P11)
\end{quote}

From the researcher's perspective, people's subtle movements were reminiscent of a form of \emph{embodied musicality}, in which participants \emph{composed} an interpretation of their experience and emotions through motion and stillness. To elaborate, the accelerated speed of agitated motion or the extended periods of lethargic thinking became the \emph{tempo} of the experience. The steady \emph{rhythmic pacing} of thought and motion, the \emph{discordance} that came from rapid motion changes, and the \emph{dynamics} of both expansive posture contrasted and slumped pose, all came together to extend this metaphor. Notably, also similar to music, these abstract expressive interpretations could potentially remain unclear and not easy to interpret by others, or sometimes even to themselves when reviewing them later.

\subsection{Interpretation of Journaled Expression}

The previous section illustrated how participants expressed themselves in embodied journaling. Here, we outline how participants reflect on and interpret their behaviours (\textbf{RQ1}) based on the affordances of each modality. 

\subsubsection{The Filter of Language}
\label{sec:expression}

As alluded to in the prior section, the way that participants expressed their experiences in the first session differed greatly in writing and in VR. Writing was seen as more precise, since words have strong meanings tied to them --- the vocabulary of language is more strongly established than interpretative motion. However, to select the precise words, writing also involved some level of cognitive processing in translation. Participants used their afforded \emph{time} to select the appropriate words they wanted to use to describe their negative experience:

\begin{quote}
    \emph{``[For writing] I think I find the right words, my word choice is more precise''} - P17
\end{quote}

\begin{quote}
    \emph{``When I'm writing, I'm choosing my words more carefully... I want it to be more coherent''} - P4
\end{quote}

As a result, writing entails a stronger \emph{filter} on emotional expression, seeing as participants needed to find and choose how to express themselves. While some participants enjoyed the process of picking the right words (e.g. P12 compared it to poetry, in which poets often find the right word to express a broad theme), others found that their linguistic vocabulary was not always expansive enough to represent how they felt. For example, P13 mentioned that \emph{``I had a hard time thinking about which words to type''}, and P6 indicated \emph{``in writing, you are bound to the language''}. 

In contrast, embodied journaling in VR did not entail a strong vocabulary for expression --- given the lack of norms that participants expressed in motion, they did not feel the same level of need to process their motion as they would for their words. P16 mentions that \emph{``[In the embodied journaling] I'm not focused on what I'm saying or how I'm phrasing things... I was just thinking''}, where thinking on the experience gives rise to natural expression rather than one tied to linguistic norms. Even though speech is also subject to language-based limitations, participants described how talking was also more unprocessed and honest compared to writing. 

\begin{quote}
    \emph{``[Embodied journaling] lets you say what's on your mind directly... [in writing], I was trying to write as well as I could.''} - P13
\end{quote}


\subsubsection{(Emotional) Connection to Self}
\label{sec:connection}

Participants shared having a strong sense of continuity with their written journals, feeling strongly connected with the textual representation of themselves --- participants generally felt that the person who wrote the text was the same as themselves in the moment. In contrast, many participants expressed a sense of disconnection with the avatar representation of themselves. Even though they were aware that the embodied avatar was expressing their motion and speech, several participants described how reviewing their embodied journal felt more akin to being a third-party observer rather than feeling like the avatar was actually themselves. P16 expressed that:

\begin{quote}
    \emph{``I watch myself as an observer --- I'm standing at a third perspective and looking at a past self... It's like two copies of myself here, and one is watching the other.''} --- P16
\end{quote}

P14 hypothesised that a potential cause for not relating strongly to their avatar was due to the disconnection between emotional expression and the interpretation required to understand themselves:

\begin{quote}
    \emph{``Because I was trying to interpret each action, I felt like... some third person watching a character in a cartoon, and that character is doing all these weird actions. It just didn't relate to me.''} --- P14
\end{quote}

Despite this, several participants \emph{empathised} with their avatar representation, resulting in emotional responses during review. Such emotional resonance was often observed when participants saw their avatar expressing unconscious cues (such as their head hanging down) that relate to feelings, rather than thoughts. 

\begin{quote}
    \emph{``I see a restricted body. I'm a very empathetic person... and that moment in time and that body posture can make me feel some emotions which I was feeling just by looking.''} - P5
\end{quote}

Moreover, some participants demonstrated a deeper understanding of the \emph{content} of their written reflections, often \emph{recalling} the reasons and meanings behind what they wrote. For example, P19 said that rereading their writing brought them back to \emph{``my realization, reflection, and growth''} from the experience. In contrast, participants noticed \emph{content-agnostic} emotional cues in their unconscious movements and expressions in embodied journaling, which they \emph{interpreted} while watching themselves. 

Noticing these emotions during the review session supported self-understanding; e.g. P4 mentioned how their \emph{``monotone''} voice reflected feeling \emph{``defeated''} and that moving in VR in general reflected their \emph{``uncomfortableness''} about the negative event, which they attempted to hide in their writing; P9 mentioned feeling \emph{``ashamed''} regarding their accusatory gestures in VR and a stronger sense of guilt about wrongdoing. Although this strong emotional resonance can support emotional discovery positively, P5 described potential danger when explored without safeguards, because \emph{``I was feeling more triggered than [the written journal]''}. This emotional intensity can create rumination, which we return to in later sections. 

\subsection{Shifted Reflective Dimensions}

Building on how participants expressed and interpreted themselves, we qualitatively compare the two journaling modalities in terms of expression, recollection, and reflection (\textbf{RQ2}).

\subsubsection{Reflection-in-Action Versus Reflection-on-Action}
\label{sec:reflectiontime}

There was a stark difference in the timing of participants' \emph{reflective} processing between the written and embodied journaling conditions. In the written condition, participants engaged in the majority of their reflective work during the creation of their written journal, consistent with the concept of \emph{reflection-in-action} \cite{yanowWhatReflectionInActionPhenomenological2009, SchonThePractitioner}. 
Conversely, in the embodied journaling condition, participants prioritised emotional expression during the first session, deferring reflective processing to the subsequent session in which they reviewed their embodied performance, which relates to the reflective practice of \emph{reflection-on-action} \cite{yanowWhatReflectionInActionPhenomenological2009}. For instance:


\begin{quote}
    \emph{``I was writing out thoughts that were more fully processed... I think that the writing was more analytical or something, like this was how I was thinking, and I think this is why... I'm not sure I got to that in VR''} - P7
\end{quote}

Several factors seemed to contribute to the temporal shift of reflection, related to the cognitive processing discussed in \autoref{sec:expression}). Participants highlighted that several affordances of written journaling facilitated reflection-in-action, including the slower pace of writing (e.g. \emph{``The pace is slower... slower means you can think deeper''} (P6)), the ability to structure thoughts on the page (e.g. \emph{``when you write things, you organise them and you make sense of every bit of it''} (P16)), and the visual feedback of seeing their written text to maintain narrative coherence (e.g. \emph{``you can see what you already wrote, so it helps you stay on top of trying to recount something from start to finish''} (P3)). 

For embodied journaling, participants remarked that analysing their body language and auditory remarks during the review session helped to explore and make connections about themselves and their emotions with fewer preconceptions. As a result, embodied journaling shifted the temporality to expressing first and reflection after the experience. Two participants reflected:

\begin{quote}
    \emph{``I think I was learning more about myself when I was watching myself.... I feel like I noticed my body language is very open... looking around very inquisitive and curious. I feel like I was learning things about myself. I feel like it was enlightening.''} - P12
\end{quote}

\begin{quote}
    \emph{``Not having specific notes about what I was thinking, allowed me freedom to come at it from a personal level... that was super helpful in just analyzing it from an external perspective and rewriting my own story about the scenario to be more compassionate to myself.''} - P8
\end{quote}

\subsubsection{Different Journaling Goals}
\label{sec:goals}
Participants reported a shift in their \emph{goals} for each journaling condition. P6 mentioned that written journaling had a strong sense of \emph{``finality''}, and many participants used writing to release their negative emotions, i.e., \emph{``when I write it down, it feels like I got that emotion out of my system''} (P12). 

However, participants sometimes realised that what they wrote did not fully align with their real emotions, which were better conveyed by their motion and vocal tone in embodied journaling. P7 expressed that disalignment as follows:

\begin{quote}
    \emph{``[In embodied journaling] I felt like I could see that I was uncomfortable in a way that I couldn't tell as much with writing... When I was writing it, I felt like I was more able to rationalise it.''} - P7
\end{quote}

Returning to the idea that people can only write about what they have thought about, P5 mentioned that \emph{``writing is better when I have a solid, complete idea, and I have understood the emotions''}, which helped people to \textit{process the whole narrative}. In contrast, participants mentioned how embodied journaling --- by making them sit through uncomfortable moments and being more abstract --- helped them understand and \emph{reveal emotions when they were not sure of how they felt}, as expressed with this quote:

\begin{quote}
    \emph{``Writing comes normally when I have an idea about what I'm feeling... [embodied journaling] was more reflective of the abstract states that I was not able to attach meaning to [through writing]''} - P5
\end{quote}

Another participant realised that they had painted themselves in a positive light in writing, but that embodied journaling helped them realise they actually felt guilty about that negative event:

\begin{quote}
    \emph{``I think unconsciously, when I write, I wrap myself in a very positive way... but it wasn't the case when I did the VR... It felt negative, but also raw; it felt more like me. (...) The gestures about blaming my partner... seeing that made me feel regret.... [In VR] I realised I had done something wrong.''} - P9
\end{quote}

Similarly, P7 realised that the conflict they previously believed to be resolved was actually still unresolved when reviewing their embodied journals, as it \emph{``made me realise that maybe I still do have some more processing to do''} (P7). Although this required sitting with the negative experiences again, P2 mentioned that \emph{``journaling is supposed to make you feel bad, if you did a bad thing. You should feel guilt and shame, because otherwise you won't change''}. 

\subsubsection{Challenges of Expressing and Interpreting Through Embodied Journaling}
\label{sec:challenges}

Even though we found that embodied journaling had reflective potential in a way that differed from written journaling, we find it imperative to highlight that it did not strongly resonate with all participants. Comfort, confidence, and emotional resonance fell on a spectrum, and the process of developing embodied journals was different for each person. Still, we do not view any of the embodied journals as being necessarily better or worse than the other; rather that they all contributed to our exploratory understanding of how the concept of embodied journaling offered insight or imagination on shaping a representation of experience. 

Even though all participants tried to do \emph{something} in embodied journaling, participants sometimes felt stuck on what to do initially as highlighted with these quotes: \emph{``I had no idea, I tried to think about what happened to try to reflect... but even then I was kind of staring at myself and looking at my VR hands''} (P20), or \emph{``I feel like I should have been doing more... it's difficult to figure out how to physically invoke any type of a living experience''} (P15).

Further, some participants reported not being able to understand and interpret what they had previously wanted to express when reviewing their avatar in session 2. For instance, P14, who re-enacted the negative experience through their actions, felt \emph{``confused, because I can see the actions, but I can't remember why the actions... reading was more clear for me''}, and P15 mentioned that \emph{``it didn't really bring up any memories in terms of the experience''}.

Even though most participants indicated that they found embodied journaling valuable despite ambiguity in representation and interpretation, a small number of participants remarked that they did not perceive its value for their own reflective purposes. The interpretation of embodied journaling was more subjective than written journaling, which may lead to different challenges in reflection and understanding. P18 mentioned that, for others viewing their embodied journal, \emph{``it's much harder for [other] people to understand my own expressions''}, but extolled its positive expressive value because \emph{``I know myself''}.

\begin{table*}[h]
    \centering
    \caption{Full table of summary statistics and Wilcoxon test values for embodied versus written journaling conditions, with n=20. The star symbol (*) denotes significant results.}
    \label{tab:stats}
    \begin{tabular}{|c|c|c|c|c|c|c|}
        \hline
        \multirow{2}{*}{\textbf{Measure}} & 
        \multirow{2}{*}{\textbf{Metric}} & 
        \multicolumn{2}{c|}{\textbf{Condition (M $\pm$ SD)}} & 
        \multirow{2}{*}{\textbf{W}} & 
        \multirow{2}{*}{$p$} & 
        \multirow{2}{*}{$r$} \\ \cline{3-4}
         &  & Embodied & Writing & & & \\ \hline
         
        \multirow{2}{*}{PANAS $\Delta$} 
            & Positive Affect & $-0.05 \pm 3.53$ & $0.75 \pm 3.89$ & 47.0 & 0.274 & -0.48 \\ \cline{2-7}
            & Negative Affect & $-1.05 \pm 3.89$ & $-1.75 \pm 3.43$ & 80.5 & 0.915 & 0.21 \\ \hline
         
        \multirow{3}{*}{R2T2 Session 1} 
            & Reflection & $9.15 \pm 2.30$ & $11.25 \pm 2.38$ & 38.5 & 0.022* & -0.51 \\ \cline{2-7}
            & Rumination & $7.35 \pm 2.39$ & $8.55 \pm 3.03$ & 41.0 & 0.160 & -0.31 \\ \cline{2-7}
            & Total (THK) & $25.50 \pm 6.40$ & $30.25 \pm 5.68$ & 39.5 & 0.014* & -0.55 \\ \hline

        \multirow{3}{*}{R2T2 Session 2} 
            & Reflection & $11.15 \pm 1.93$ & $10.70 \pm 2.08$ & 55.0 & 0.496 & 0.15 \\ \cline{2-7}
            & Rumination & $9.30 \pm 2.23$ & $8.30 \pm 2.64$ & 47.5 & 0.168 & 0.31 \\ \cline{2-7}
            & Total (THK) & $30.60 \pm 5.59$ & $30.50 \pm 4.65$ & 82.5 & 0.614 & 0.11 \\ \hline

        \multirow{10}{*}{MEQ} 
            & Vividness & $11.95 \pm 2.52$ & $12.05 \pm 2.65$ & 43.5 & 0.888 & -0.03 \\ \cline{2-7}
            & Coherence & $16.65 \pm 3.10$ & $16.60 \pm 3.30$ & 47.0 & 0.727 & 0.08 \\ \cline{2-7}
            & Accessibility & $13.35 \pm 2.08$ & $13.40 \pm 2.58$ & 43.5 & 0.887 & -0.03 \\ \cline{2-7}
            & Time Perspective & $11.40 \pm 2.48$ & $10.50 \pm 2.69$ & 9.0 & 0.058 & 0.42 \\ \cline{2-7}
            & Sensory Details & $13.75 \pm 2.51$ & $13.10 \pm 2.92$ & 50.0 & 0.205 & 0.28 \\ \cline{2-7}
            & Visual Perspective & $11.80 \pm 2.76$ & $13.15 \pm 2.74$ & 33.0 & 0.066 & -0.41 \\ \cline{2-7}
            & Emotional Intensity & $10.70 \pm 2.75$ & $10.25 \pm 2.61$ & 15.5 & 0.215 & 0.28 \\ \cline{2-7}
            & Sharing & $9.30 \pm 3.33$ & $9.00 \pm 3.13$ & 6.0 & 0.330 & 0.22 \\ \cline{2-7}
            & Distancing & $9.15 \pm 3.03$ & $8.95 \pm 2.86$ & 20.0 & 0.765 & 0.07 \\ \cline{2-7}
            & Valence & $5.60 \pm 1.57$ & $5.60 \pm 1.73$ & 10.5 & 1.000 & 0.00 \\ \hline
    \end{tabular}
    \Description{This table shows the mean and standard deviation for each of the metrics we evaluated for both embodied and written journaling conditions. It also shows the Wilcoxon statistic (W), the p-values, and the effect size (r).}
\end{table*}

\section{Quantitative Findings}\label{sec:quan}

To consider changes in affect during journal creation (in the first session), we first calculated the positive and negative affects of the PANAS before and after each condition. We then took the difference between pre- and post-condition to compare the net change in positive and negative affect across the two conditions. In both conditions, there was no meaningful change in positive affect, while negative affect seemed to decrease slightly. Comparing across conditions, written and embodied journaling showed no significant difference in the changes of both positive (\emph{W}=47.0, \emph{p}=.274) and negative affect (\emph{W}=80.5, \emph{p}=.915), suggesting that neither modality drove mood shifts directly after creation.

\begin{figure*}[h]
  \centering
  \includegraphics[width=0.7\linewidth]{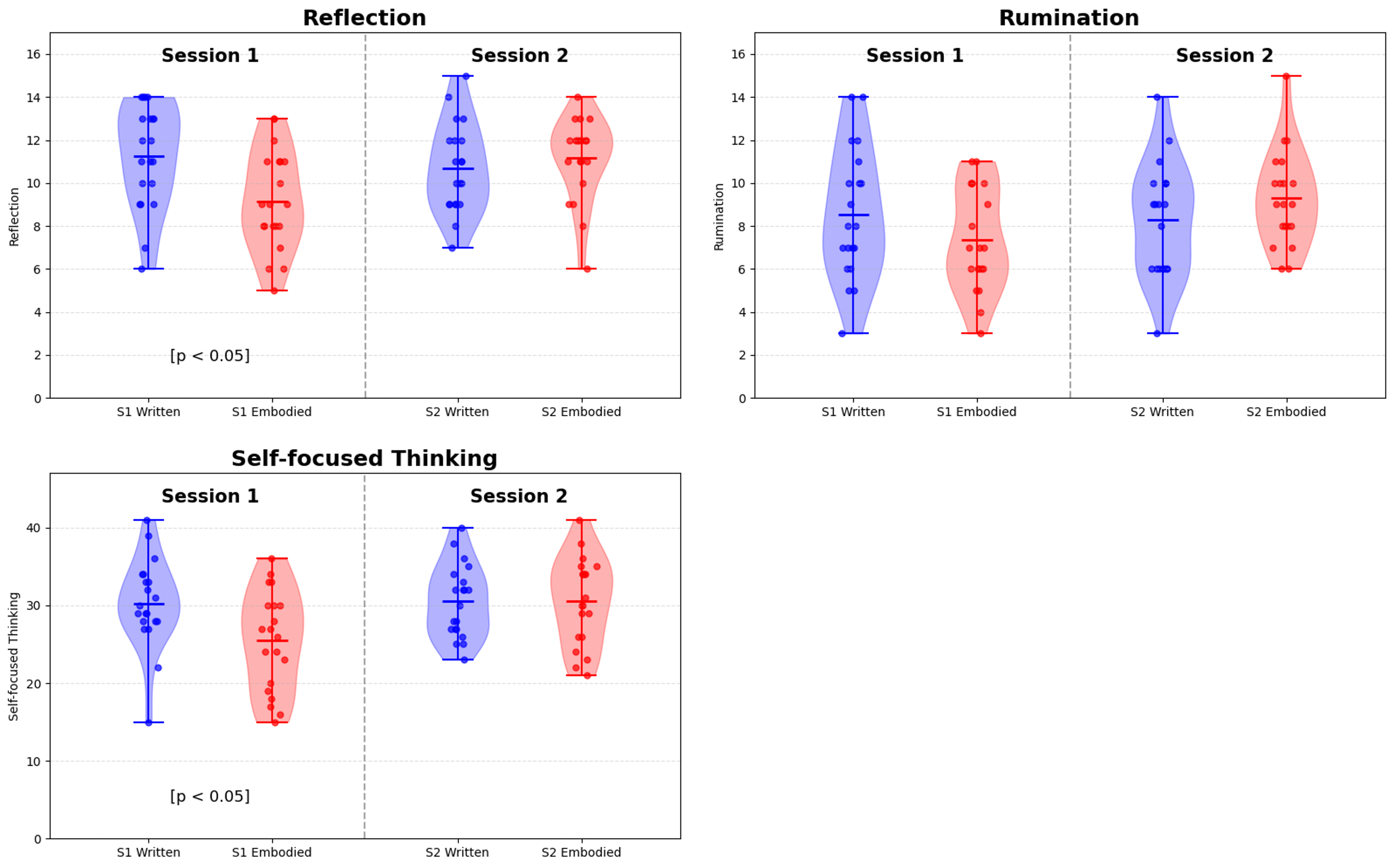}
  \caption{Violin plots for each of the metrics of the R2T2 scale \cite{loerakkerTechnologyWhichMakes2024b}, with line markers for the max, mean, and min. }
  \Description{This image shows 3 violin plots for the metrics of the R2T2 scale of reflection, rumination, and self-focused thinking. There are significant differences in the embodied and VR conditions in the first session for the reflection and self-focused thinking metrics. }
  \label{fig:r2t2}
\end{figure*}

We examined the R2T2's reflection subscale, rumination subscale, and composite self-focused thinking score across both sessions \cite{loerakkerTechnologyWhichMakes2024b} --- see Figure \ref{fig:r2t2}. After the first session, we found that values for all three measures were higher for writing over embodied journaling. In particular, there were statistically significant differences in reflection (\emph{W}=41.0, \emph{p}=.022) and self-focused thinking (\emph{W}=39.5, \emph{p}=.014). This complements our qualitative findings --- reflective processes were more prominent during writing (as a form of \emph{reflection-in-action}). However, when we asked participants the same scale in the second session, these statistically significant differences were no longer apparent. Not only that, but all three metrics were higher for embodied journaling over written journaling. This complete reversal aligns strongly with the temporal discussion of our qualitative findings, in which reflective processes come from review of embodied journaling (as a form of \emph{reflection-on-action}). This is further supported by the ART ANOVA that compared across measure and session, as all three measures of reflection (\emph{F}=8.76, \emph{p}=.004), rumination (\emph{F}=4.81, \emph{p}=.032), and self-focused thinking (\emph{F}=4.55, \emph{p}=.037) reported statistically significant differences, highlighting how all three reflective measures change between the two sessions. 

Finally, in terms of memory recollection, we found that none of the metrics for memory experience recall were strongly significant, especially when taking into account the multiple comparisons correction. However, the quantitative data still suggests some looser trends in time perspective (\emph{W}=9.0, \emph{p}=.058) and visual perspective (\emph{W}=33.0, \emph{p}=.066). This temporal perspective aligns with participant responses, in which participants reported that embodied journaling helped capture a more concrete moment, whereas written journaling captured the entirety of the narrative. The possible trend in visual perspective also aligns with the responses, in which participants often recounted perceiving themselves as an external observer rather than from their own eyes in embodied journaling.  

Although we counterbalanced for the order of embodied and written journaling across the two sessions in our work, we acknowledge that the two conditions are not independent and can have quantitative carryover effects \cite{hoCausalInferenceCounterbalanced2025}. When participants are invited to reflect on an experience once already, this may affect how they reflect on the same experience a subsequent time. We found no significant carryover effects, except for one dimension — the change in positive affect in embodied journaling in the first session (\emph{p} = .024). In particular, the positive affect showed a relative increase for embodied journaling when it was performed first (and a slight decrease when performed second). This potentially ties back to the temporal trend of reflection --- in that when embodied journaling was performed after written journaling, the reflective labour had already been done.

\section{Discussion}
Reflection has become increasingly important in the HCI literature, as it helps people communicate, question, and make sense of their world and their experiences. We explored how embodied journaling --- using physical motion and voice --- influences emotional processing, reflection, and recollection, and how the experiences and effects contrast with traditional written journaling. While we formulate implications and provide suggestions for the future based on these findings, their exploratory nature means they should be understood to be context-dependent and conditional. 

\subsection{Expression in Different Forms}

In embodied journaling, the body becomes the site for experiential expression. In written journaling, thought, language, and writing are entangled, which means that one can only express what has been consciously thought of. In embodied journaling, motion and paralinguistic speech cues partially emerged from the unconscious, representing feelings that were more raw and honest (as in Section \ref{sec:goals}). By unbinding expression from linguistic meaning, participants reported expressing feelings that were even unconscious to themselves. This strongly relates to somaesthetic design, where the body becomes a site for understanding, cognition, and self-expression \cite{shustermanBodyArtsNeed2012, shustermanSomaestheticsDisciplinaryProposal1999} and movement emerges \emph{naturally} to create meaning \cite{hookDesigningBodySomaesthetic2018}. This natural usage of body movement to create meaning contrasts to intentionally using the body as a way to drive meaning-making of oneself and the world \cite{mauMentalMovementsHow2021, hirschMyPaceProvokes2011, buttingsrudBodiesSkilledPerformance2021}, as is particularly prominent in dancing \cite{buttingsrudBodiesSkilledPerformance2021, schwenderCreatingFramingReflecting2025}.

Intentional and conscious motion are important --- the broad patterns of motion are connected with thinking and reflection \cite{mauMentalMovementsHow2021, hirschMyPaceProvokes2011, buttingsrudBodiesSkilledPerformance2021}. \citet{buttingsrudBodiesSkilledPerformance2021} and \citet{schwenderCreatingFramingReflecting2025} outline the experience of embodied reflection through dance as a way to drive meaning-making of oneself and the world. However, for our participants, who lacked the strong intentional vocabulary of movement associated with dancers, it was generally the retrospective \emph{review} of motion that promoted deeper reflection. The cues that participants were surprised by during embodied journaling were subtle and microscopic --- head motion, swaying, and fidgeting. Unlike choreographed movements, these motions invited participants to ponder \emph{why} their body exhibited such cues. In doing so, they were able to learn more about their \emph{primary} feelings towards the experience. Even for spoken voice, which does abide by linguistic boundaries, it was the oft-unconscious paralinguistic cues, such as corrections, phrasing, and tone \cite{guyerParalinguisticFeaturesCommunicated2021}, that were representative of emotions; cues often lost in written journals. While language plays an important role in consciousness and self-awareness \cite{carruthersCognitiveFunctionsLanguage2002}, reflection does not necessarily need to end with linguistic reasoning. 

\textbf{\textsc{Recommendation 1}\textemdash } Consequently, we suggest that reflective systems may incorporate bodily and paralinguistic expression, potentially enabling users to access unconscious feelings during moments of self-reflection.  

\subsection{User Connection and Embodiment}

Even though participants resonated with the embodied avatar, they simultaneously reported a personal disconnection from their previous self in Section \ref{sec:connection} --- rather than re-inhabiting the avatar with their motion and speech, they perceived it through the lens of an external observer. We hypothesise that the previously mentioned unconscious embodied expression partially contributed to these feelings of disconnection. We draw from construal level theory (CLT) \cite{tropeConstruallevelTheoryPsychological2010}, which informs human construal formation --- written journaling was closer to the person's thoughts, leading to concrete, specific, and strongly connected construals; embodied journaling felt further from their thoughts, leading to broad, abstract, and disconnected construals. This disconnection could potentially lead to higher cognitive load in both creation and review, as more mental load is needed to interpret what was represented and why (see Section \ref{sec:challenges}). However, this could also tie into the cognitive dissociation inherent to out-of-body experiences in VR \cite{maselliSlidingPerspectivesDissociating2014, bourdinVirtualOutofBodyExperience2017,vanheugten-vanderkloetOutofbodyExperienceVirtual2018a} --- more work is needed to explore this potential effect. 

Reflective VR systems often use interactive elements or prompted tasks to guide reflection \cite{yinTravelGalleriaSupportingRemembrance2025, wagenerSelVReflectGuidedVR2023a, bahngDesigningImmersiveStories2023a}, which lightens this cognitive load. We stripped such elements in our work to understand at a fundamental level what reflective tendencies and challenges are afforded through solely motion and speech, yet doing so also possibly created difficulty in participants' understanding what to do, and thus, how to interpret what they did (Section \ref{sec:challenges}). Future work can once again reintroduce interactive digital artifacts or other key features to support and guide users' diverse reflective needs. Inspired by Jiang and Ahmadpour's design suggestions for VR \cite{jiangImmersionDesigningReflection2022a}, some ideas could be to construct a \emph{personally relevant setting} instead of a minimalistic virtual space, or using \emph{personalised artifacts} and annotation tools to add explanability to raw motion and speech. Such features may bridge the gap between bodily expression and the user’s intended goals, making the embodied journal feel more understandable and personally meaningful. 

Embodiment is strong in VR \cite{schultzeEmbodimentPresenceVirtual2010b, kilteniSenseEmbodimentVirtual2012c}, and avatars typically serve as an expression of the user. Yet, participants felt separated from their movement, even when prior work on avatarization has implied the inverse \cite{freemanMyBodyMy2020b}. We speculate that people strongly resonate with avatar properties that they have control over and can choose, such as their appearance, aesthetics, and overall self-presentation \cite{freeman2021body}. In contrast, movement, which is much more primary and uncontrollable, disconnects people from their desired self-representation. Overall, this reveals a complexity in embodiment theory, potentially highlighting a tension between control and exposure. Ultimately, there exist mediating factors of reflection centred around the design of the VR system, including the avatar and its self-representation and the world and its interactive elements.

\textbf{\textsc{Recommendation 2}\textemdash } Thus, we recommend that future systems that facilitate embodied journaling in VR could explore participant support through intentional design and interactive elements to encourage expression, interpretation, and customization.

\subsection{The Temporality of Reflection}

In our study, people's written journals became sites of narrative creation, self-inquiry, and understanding, becoming \emph{reflection-in-action} \cite{SchonThePractitioner}. The process of writing, when compared to embodied journal creation, was more reflective on both qualitative and quantitative levels. Under Fleck and Fitzpatrick's framework about the levels of reflection, participants were able to engage up to the level of transformative thinking \cite{fleckReflectingReflectionFraming2010}; under Baumer's reflective dimensions, participants were able to break down their experiences and inquire into their motivations and feelings \cite{baumerReflectiveInformaticsConceptual2015a}. However, we also highlighted some pitfalls of writing in Section \ref{sec:goals}. For example, because thoughts had been processed, they could be perceived as more final, and this sense of finality sometimes led to emotional performativity --- self-presenting oneself at the end of a negative experience in a positive light because of what one wants to believe. 

The process of \emph{embodied creation} would not traditionally be considered deep reflection within existing reflective frameworks, as it focuses on experience recounting \cite{fleckReflectingReflectionFraming2010, baumerReflectiveInformaticsConceptual2015a}. Instead, reflection occurred when reviewing (\emph{reflection-on-action} \cite{SchonThePractitioner}), which prior work has stated can potentially extend upon the reflections during creation \cite{bolton2018reflection}. Quantitatively and qualitatively, the difference in reflection between the two mediums was much less evident after review, indicating how participants could experience a similar depth of transformative thinking \cite{fleckReflectingReflectionFraming2010} and self-inquiry \cite{baumerReflectiveInformaticsConceptual2015a}. 

While embodied and written journaling both encourage reflection through revisiting the past and surfacing memories \cite{bentvelzenRevisitingReflectionHCI2022}, their expression of memories was different --- written journaling facilitates exact narrative depictions where expressing and processing are more intertwined, whereas we observed embodied journaling as building abstract representations with a stronger focus on emotional expression during creation and processing during review. The totality of embodied journaling (creation plus review) evokes a level of implicit slowness for reflection \cite{odomExtendingTheorySlow2022a} in comparison to writing. Furthermore, reflective discovery was mediated through ambiguity in embodied journaling \cite{bentvelzenRevisitingReflectionHCI2022}, since abstract movements must be reinterpreted as feelings or experiences, contrasting written content, which tends to be more easily re-understood. As such, our exploration of embodied journaling surfaces how reflection takes place across multiple temporal stages, mediated by modality. In particular, for some participants, we noticed that reflection that takes place too quickly during writing can risk not drawing out the complete picture of self-understanding. 

\textbf{\textsc{Recommendation 3}\textemdash } Therefore, we encourage future reflective systems to view reflection as an extended temporal practice to experience different forms of discovery.

\subsection{Managing Emotional Discomfort}

Prior works have positioned written journaling as a useful practice because of its transformative quality to improve mood and well-being \cite{ullrichJournalingStressfulEvents2002c, mercerVisualJournalingIntervention2010a, richelleStayPositiveEffects2024, isikEffectsGratitudeJournaling2017a}. However, what is often missed is that growth necessitates also sitting with and accepting feelings, including negative ones \cite{shallcrossLetItBe2010}. Increasingly, journaling systems attempt to abstract away the human step of self-understanding, e.g. by using AI for sensemaking \cite{kimDiaryMateUnderstandingUser2024b, wangDesigningAIAugmentedJournaling2025a}. 

While written journaling can rapidly reframe these negative feelings into takeaways, the separation between expression and reflection of embodied journaling could make `sitting with one's feelings' more pronounced. As implied by some participants, this can come at the risk of rumination --- becoming stuck in negative feelings \cite{krossSelfReflectionWorkWhy2023, harringtonInsightRuminationSelfReflection2010, marinRuminationSelfreflectionStress2017, eikeySelfreflectionIntroducingConcept2021}. Still, sitting with one's feelings to fully embrace them is not necessarily negative in and of itself --- it is a form of self-inquiry and self-understanding \cite{shallcrossLetItBe2010, kivitySelfAcceptanceNegativeEmotions2016, fordPsychologicalHealthBenefits2018} aligned with Wagener et al.'s work, which emphasises the importance of \emph{holding} \cite{wagenerLettingItGo2023} and \emph{materializing} \cite{wagenerMoodShaper2024} negative emotions to process them into growth. 

In Section \ref{sec:goals}, we found that written journaling can help provide finality to negative emotions that one already understands; one can communicate and draw takeaways of the experience, and write down goals for future transformations \cite{bodenFivePerspectivesReflective2006}. On the other hand, embodied journaling could spotlight feelings one may not understand as an enduring process of self-inquiry. By expressing, one can sit with their feelings, put their feelings on trial, and draw takeaways after confronting themselves. Both manners of managing emotions are valid, but we find that prior work often tends towards the former - participants were able to learn from revisiting their negative experiences, even if it came with guilt or sadness. 

\textbf{\textsc{Recommendation 4}\textemdash } As a result, we suggest that future systems may benefit from exploring ways to embrace negative emotions rather than immediately seeking to expel them, while also supporting the user when they are veering into rumination. 


\section{Limitations and Future Work}

There were a few methodological limitations of our work that correspond with its exploratory nature. Starting with our sample demographics --- it leaned towards young adults who generally scored highly on the SRIS; future work could explore embodied journaling with a more diverse demographic of users, with participants who may be less familiar with introspection. Our participants had diverse experiences with VR and with journaling. For less familiar users in these domains, novelty effects may have played a role in people's expression and reflection. In addition, even though our work employed within-subject testing for exploratory comparison, future work can broaden to between-subject evaluation to reduce these modality-order effects.  

The goal of this study was to spotlight the possible \emph{range} of experiences, outcomes, and effects. However, future work can take a more confirmatory perspective to validate the \emph{effectiveness} of embodied journaling. Since our study involved only one session, which can limit the outcome of reflection \cite{pascual-leoneDoesFeelingBad2016}, longitudinal field studies could examine the extended use of embodied journaling. Further, many participants in our study expressed initial insecurity and uncertainty about how to use their bodies for journaling before gaining confidence, and their familiarity with using their bodies was varied (e.g. P5 had an informal level of dance experience). Future work can explore how patterns of using one's body for journaling might shift as people become more comfortable with embodied expression, furthering the need for a longitudinal study.

Such a study would also look at embodied journaling in-the-wild, understanding its use across diverse experiences in the moment to see how people express and reflect on themselves in the midst of intense emotional moments (rather than the more muted, resolved experiences from our study). Performing this study with a broader, more diverse population would also enable stratifications around individual details and their reflective outcomes that can induce more granular evaluation, such as how individual levels of body ownership might mediate how people reflect on their journals. 

From the technical side, we used inverse kinematics to align the avatar with the users' motions in real-time to ensure a low-cost, low-tech setup suitable for a wide range of users, fitting the use case of at-home embodied journaling. More accurate synchronous tracking could improve realism at the cost of additional sensory hardware. Furthermore, the experimental space of embodied journaling was limited due to space constraints and VR cabling. We propose increasing the physical freedom of embodied journaling through wireless interaction or larger interaction spaces. Lastly, while some participants did use improvised external props (i.e., a chair) as part of journaling, future work can consider a more structured implementations of props, which could support emotional expression \cite{stadlerEffectPropsStory2010}.

\section{Conclusion}
We explored a somaesthetic approach to journaling that utilises the human body --- and its voice and motion --- as the medium of reflection. We implemented a system in VR that supports \emph{embodied journaling}, allowing users to journal about a negative experience using the body, and then review their recording at a later time. In a comparative user study with 20 participants, we find that participants used speech, silence, motion, and stillness to form a vocabulary of embodied journaling. In comparison to written journaling, this process facilitated deeper emotional expression beyond the linguistic boundaries of words, allowing participants to understand aspects of themselves and emotions that they had been previously unaware of. We discuss that both embodied and written journaling spark reflection, albeit at different times during the process and with different positive benefits and challenges. We propose embodied journaling as a valuable approach for future research to uncover and reflect on hard-to-verbalise feelings that persist even after linguistic processing and resolution.

\begin{acks}
This work was supported in part by the Natural Science and Engineering Research Council of Canada (NSERC) under Discovery Grant RGPIN-2019-05624.
\end{acks}

\bibliographystyle{ACM-Reference-Format}
\bibliography{sample-base}

\end{document}